\documentstyle{article}

\newcounter{wsf}
\def\ncd{\newcommand}

\def\mtg{{\sf MTG}}
\def\mcl{Mac~Lane}
\def\mlm{\mcl\ method}

\ncd{\whg}{{\sf WHG}}

\ncd{\eq}{\begin{equation}}
\ncd{\qe}{\end{equation}}
\ncd{\eqas}{\begin{eqnarray*}}
\ncd{\saqe}{\end{eqnarray*}}
\ncd{\refe}[1]{(\ref{#1})}
\ncd{\nn}{\nonumber\\}

\ncd{\MOP}[2]{\ncd{#1}{\mathop{{\rm #2\,}}\nolimits}} 
\MOP{\Ker}{Ker} 
\MOP{\Aut}{Aut} 
\MOP{\Hom}{Hom} 
\MOP{\Tr}{Tr}   

\ncd{\xb}{\bar{x}} \ncd{\rb}{\bar{r}} \ncd{\fb}{\bar{f}}
\ncd{\Ab}{\bar{A}} \ncd{\Bb}{\bar{B}}
\ncd{\ovr}[1]{\overline{#1}}

\ncd{\vA}{{\bf A}} \ncd{\vH}{{\bf H}} \ncd{\vS}{{\bf S}} \ncd{\vB}{{\bf B}}
\ncd{\va}{{\bf a}} \ncd{\vh}{{\bf h}} \ncd{\vs}{{\bf s}} \ncd{\vp}{{\bf p}} 
\ncd{\vz}{{\bf0}} \ncd{\vt}{{\bf t}} \ncd{\vk}{{\bf k}} \ncd{\vr}{{\bf r}}
\ncd{\ve}{{\bf e}}
\ncd{\bi}[1]{\mbox{{\scriptsize\bf #1}}}
\ncd{\ivt}{\bi{t}} \ncd{\ivk}{\bi{k}} \ncd{\ive}{\bi{e}}
\ncd{\pib}{\mbox{\boldmath$\pi$}\!}
\ncd{\cQ}{{\cal Q}} \ncd{\cP}{{\cal P}} \ncd{\Tg}{{\cal T}}
\ncd{\vrho}{\varrho} \ncd{\vph}{\varphi} \ncd{\veps}{\varepsilon}
\ncd{\ii}{{\rm i}} \ncd{\ee}{{\rm e}}
\ncd{\uo}{{\sf U}(1)}
\ncd{\mathz}{\mbox{\sf Z\hspace{-0.90ex}Z}}
\ncd{\mathr}{\mbox{\sf l\hspace{-0.30ex}R}}
\ncd{\mathc}{\mbox{\sf C\hspace{-1.00ex}l\hspace{0.45ex}}}
\ncd{\matha}{\mbox{\sf A}}
\ncd{\mathb}{\mbox{\sf B}}
\ncd{\integ}{{\mathz}}  
\ncd{\real}{{\mathr}}   
\ncd{\compl}{{\mathc}}  
\ncd{\mata}{{\matha}}   
\ncd{\matb}{{\mathb}}   
\ncd{\cm}{\mathc^*}

\ncd{\cdt}{\!\cdot\!} \ncd{\ra}{\!\to\!} \ncd{\tims}{\!\times\!} 
\ncd{\ce}{\odot}
\ncd{\half}{{\textstyle{1\over2}}}

\ncd{\eqa}{\begin{eqnarray}}
\ncd{\aqe}{\end{eqnarray}}

\ncd{\eqax}{                     
\stepcounter{equation}\setcounter{wsf}{\value{equation}}%
\setcounter{equation}{0}%
\def\theequation{\arabic{wsf}\alph{equation}}\eqa}
\ncd{\eqaxl}[1]{                     
\refstepcounter{equation}\label{#1}\setcounter{wsf}{\value{equation}}%
\setcounter{equation}{0}%
\def\theequation{\arabic{wsf}\alph{equation}}\eqa}
\ncd{\aqex}{\aqe\setcounter{equation}{\value{wsf}}%
\let\theequation\thewsf}

\begin{document}

\title{Magnetic translation groups in $n$ dimensions\thanks{This work was
realized within the project No {\bf PB 201/P3/94} supported by the 
Polish State Committee for Scientific Research (KBN).}}
\author{W. Florek\thanks{e-mail address: florek@amu.edu.pl}\\[3mm]
Institute of Physics, A. Mickiewicz University,\\
ul. Umultowska 85, 61--614 Pozna\'n}
\date{\today} 
\maketitle

\begin{abstract}
 Magnetic translation groups are considered as central extensions of the
translation group $T\simeq\integ^n$ by the group of factors (a~gauge
group)~\uo\@. The obtained general formulae allow to consider a~magnetic
field as an~antisymmetric tensor (of rank 2) and factor systems are
determined by a~transvection of this tensor with a~tensor product
$\vt\otimes\vt'$\@. 

\end{abstract}

\section{Introduction}\label{intro}
 The behaviour of electrons in crystalline (periodic) potentials in the
presence of a constant (external) magnetic field has been studied since the
thirties in many papers, amongst which works of Landau~\cite{landau},
Peierls~\cite{peierls}, Onsager~\cite{onsager}, Harper~\cite{harper}, and
Azbel~\cite{azbel} seem to be ones of the most important. In the sixties
Brown \cite{brown} and Zak \cite{zak1,zak2} (see also \cite{zak6})
independently introduced and investigated the so-called {\em magnetic
translation groups} (\mtg)\@.\footnote{The term {\em `magnetic translation 
group'} the first time was used by \mbox{Peterson}~\cite{peter}.} Their
results have been lately applied to a~problem of the quantum Hall
effect~\cite{dana1,dana2} and relations with the Weyl--Heisenberg group
(\whg) have been also studied~\cite{zak}\@. Some interesting results have
been presented lately by Geyler and Popov~\cite{geypop}\@.

From the group-theoretical point of view magnetic translations can be
considered as a projective (ray) representation of the translation group~$T$
of a~crystal lattice (this is Brown's approach)\@. However, projective
representations of any group can be found as vector representations of its
covering group, which can be constructed as a~central extension of a~given
group by the group of factors being, in general, a~subgroup of~$\cm$ (see,
e.g., \cite{barut,alt})\@. This construction is a~basis of Zak's
considerations and it is also used in this paper, since obtained results
allow us to see more general properties of \mtg\ and we can get a~deep
insight in its (algebraic) structure.

The aim of this work is to investigate \mtg\ for a~$n$-dimensional crystal
lattice, i.e.\ to investigate central extensions of $\integ^n$ by \uo\@.
Such approach leads to interpretation of a~magnetic field as an
antisymmetric $n$-dimensional tensor of rank~2 and to determine mathematical
background of flux quantization. All nonequivalent central extensions of 
$\integ^n$ by \uo\ are found by the \mlm\ for determination of the second
cohomology group~\cite{mlm1,mlm2} (see also \cite{mll} and references quoted
therein), which becomes significantly simplified in the considered case of
a~central extension of two abellian groups. The important (especially from
physical point of view) problem of determination of irreducible
representations can be solved using the induction procedure (see, e.g.,
\cite{alt}) but it is left over to further considerations. Nevertheless, the
obtained results can be compared with those of Brown and Zak and, therefore,
one can find physical meaning of introduced parameters.

The facts indicating that \mtg\ can be considered as a~central extension are
presented in Sec.~\ref{cemtg}. The relations with \whg\ are also pointed
out. In Sec.~\ref{ext} all non-equivalent central extensions of $\integ^n$
by \uo\ are determined. Moreover, a~labelling scheme for obtained extensions
is introduced and its physical relevance is indicated. Some crucial, but
cumbersome, calculations can be found in the appendix. As an example the case
$n\!=\!3$ is considered in Sec.~\ref{ext3} and the results is compared with
Brown's and Zak's ones.

\section{\mtg, \whg\ and Central Extensions} \label{cemtg}

The Weyl--Heisenberg group is generated by unitary operators ($a$ and $b$ are
real constants)
 \eq \label{QP}
  \cQ(a)=\exp(-\ii aQ), \qquad \cP(b)=\exp(-\ii bP),
 \qe
 where~$(Q,P)$ is a~pair of complementary (hermitian) operators, i.e.
 \eq \label{comm}
  [P,Q]=-\ii\hbar,
 \qe
 The operators~$\cQ$ and~$\cP$ satisfy the following relation
\cite{weyl,thir} (see also~\cite{schw}, where a~finite phase plane is
discussed)
 \eq\label{commQP}
  \ee^{-\ii bP} \ee^{-\ii aQ} = 
   \ee^{-\ii aQ}\ee^{-\ii bP}\ee^{\ii\hbar ab}.
 \qe
 In the case of \mtg's the r\^oles of~$Q$ and $P$ are played by components,
$\pi^c_x$ and $\pi^c_y$ respectively, of the vector operator (cf.\
\cite{zak})
 \eq
 \pib^{\,c}=\vp-\frac{e}{c}\vA
 \qe  
 (which has a meaning of the center of the Landau orbit \cite{JL}), where
\vA~is the vector potential of the magnetic field\footnote{The gauge
$\vA=(\vH\times\vr)/2$ is used in this work.} $\vH$ and \vp~is the momentum
operator. It is easy to check that for $\vH\!=\![0,0,H]$
 \eq\label{commpi}
  [\pi^c_y,\pi^c_x]=-\ii\hbar \frac{eH}{c}
 \qe
 and substituting
 \eq\label{QPtopi}
 Q=\frac{c}{eH}\pi^c_x,\qquad P=\pi^c_y
 \qe
 the relation~\refe{comm} is revived. Magnetic translation operators (for
the sake of simplicity the square lattice, determined by orthogonal vectors
$\va_x\!=\![a,0]$ and~$\va_y\!=\![0,a]$, is considered) can be introduced as
 \eq
    T(n_\xi\va_\xi)=\exp(-\ii n_\xi a\pi^c_\xi/\hbar), \qquad \xi=x,y.
 \qe
 Using the formulae~\refe{QP} and~\refe{QPtopi} one can also write
 \eqax
    T(n_x\va_x)&=&\ee^{-\ii n_x a eHQ/c\hbar}=\cQ(n_x aeH/c\hbar);\\
    T(n_y\va_y)&=&\ee^{-\ii n_y a P/\hbar}=\cP(n_y a/\hbar).
 \aqex
 Therefore, the commutation rule \refe{commQP} yields
 \eqa
  T(n_x\va_x)T(n_y\va_y)&=&T(n_y\va_y)T(n_x\va_x)
   \exp(-\ii n_x n_y a^2 eH/c\hbar)\nn[3mm]
   &=&T(n_y\va_y)T(n_x\va_x)
   \exp[-\ii(n_x\va_x\tims n_y\va_y)\cdt\vH e/c\hbar].
 \aqe
 Since $[\pi^c_y,\pi^c_x]$ is a~complex number, the relation
 $$
  \ee^A\ee^B=\ee^{A+B+[A,B]/2}
 $$
 can be used and for $\vt\!=\![n_xa,n_ya]$ one obtains 
 \eq
  T(\vt)=\exp(-\ii\vt\cdt\pib^{\,c}/\hbar)=
  T(n_x\va_x)T(n_y\va_y)\exp[\half\ii n_xn_ya^2He/c\hbar].
 \qe
  This formula defines a projective representation of a~(two-dimensional)
translation group, which in the absence of a~magnetic field reduces to
a~pure translation operator acting on a~function~$\psi(\vr)$ in a~standard 
way, i.e.\ (cf.\ also \cite{brown} and \cite[Chaps.\ 1 and 8]{alt})
 \eq\label{proj}
  T(\vt)|_{\vA=\vz}=\exp(-\ii\vt\cdt\vp/\hbar)\quad
  {\rm and}\quad T(\vt)|_{\vA=\vz}\psi(\vr)=\psi(\vr\!-\!\vt).
 \qe

As it was mentioned in Sec.~\ref{intro} consideration of projective
representations can be replaced by investigation of a~covering group and its
vector representattions \cite{barut,alt}\@. Zak~\cite{zak1,zak2,zak} defined
a~covering group~$\Tg$ of the translation group~$T$ as a~set of the
following operators\footnote{Note that in the presented definition the sign
is changed as compared with Zak's paper~\cite{zak1} in order to obtain later
equations consitent with \refe{proj} and, moreover, with other works of Zak
(e.g.~\cite{zak}).}
 \eq \label{zakd}
  \tau(\vt\mid\vt_1,\vt_2,\ldots,\vt_l)=
  \exp(-\ii\vt\cdt\pib^{\,c}/\hbar)
  \exp[-\ii e\Phi(\vt_1,\vt_2,\ldots,\vt_l)/c\hbar],
\qe
 where~$\vt$ is a~lattice vector, $\vt_1,\vt_2,\ldots,\vt_l$ is any path
joining the origin~$O$ with the point defined by~$\vt$ (i.e.\
$\vt\!=\!\sum_{j=1}^l\vt_j$; all $\vt_j$, $1\!\le\!j\!\le\!l$, are lattice
vectors), and $\Phi(\vt_1,\vt_2,\ldots,\vt_l)$ is the magnetic flux through
the polygon enclosed by the vectors $\vt_1,\ldots,\vt_l,-\vt$, i.e.\
$\Phi\!=\!\vH\cdt\vS$, with \vS\ being the area of the mentioned polygon. 
This area can be calculated as (see \cite{zak1}) 
 \eq
  \vS=\half(\vt_1\tims\vt_2+\ldots+\vt_1\tims\vt_l+\vt_2\tims\vt_3+
  \ldots+\vt_{l-1}\tims\vt_l), 
 \qe
 i.e.\ by calculating all ${n\choose2}$ vector products (in the order
determined by the order of vectors in the path)\@. 

On the other hand the Hamiltonin for an electron in a periodic potential
$V(\vr)$ and a uniform magnetic field (described by the vector potential
\vA) is given as \cite{brown,zak1}
 \eq
  {\cal H}=\frac{1}{2m}\pib^{\,2}+V(\vr),
 \qe
 where   
 \eq 
  \pib=\vp+\frac{e}{c}\vA 
 \qe
 is the (vector) operator of the kinetic momentum. The operators defined in
\refe{zakd} commute with this Hamiltonian if the vector potential~\vA\
fulfils the condition
 \eq 
  \partial A_\xi/\partial \chi+\partial A_\chi/\partial \xi=0;\qquad 
   {\rm for\ }  \xi,\chi=x,y,z.
\qe
 This relation holds, for example, for the gauge $\vA\!=\!(\vH\tims\vr)/2$, 
which was used by both authors \cite{brown,zak1} and will be applied in this
work. 

In his first paper Zak showed that
 \eqa
\tau(\vt\mid\vt_1,\ldots,\vt_l)
\tau(\vt'\mid\vt'_1,\ldots,\vt'_j) &=&
\tau(\vt\!+\!\vt'\mid
  \vt_1,\ldots,\vt_l,\vt'_1,\ldots,\vt'_j);\\
\tau(\vt\mid\vt_1,\ldots,\vt_l) \tau(\vt'\mid\vt'_1,\ldots,\vt'_j) &=&
\tau(\vt'\mid\vt'_1,\ldots,\vt'_j) 
\tau(\vt\mid\vt_1,\ldots,\vt_l)\nonumber.\\ &&
\tims\exp\left[-\ii e(\vt\tims\vt')\cdt\vH/c\hbar\right]. 
 \aqe
 These relations yield that elements $(\vz\mid\vt_1,\ldots,\vt_l)$ (with
$\sum_{j=1}^l\vt_j\!=\!\vz$) belong to the center $Z(\Tg)$\@. Moreover, the
quotient group $\Tg/Z(\Tg)$ is isomorphic with the group~$T$ of (ordinary)
translations~\cite{zak1}, and one can choose as representatives of {\em
right} cosets elements $\tau(\vt\mid\vt)$\@. It is easy to show that for any
lattice vector~\vt\ the operators $\tau(\vt\mid\vt)$ and
$\tau(\vt\mid\vt_1,\vt_2)$, where $\vt_1\!+\!\vt_2\!=\!\vt$ belong to the
same right coset
 \eq\label{dec}
  \tau(\vz\mid\vt_1,\vt_2,-\vt)\tau(\vt\mid\vt)=
  \tau(\vt\mid\vt_1,\vt_2,-\vt,\vt)=\tau(\vt\mid\vt_1,\vt_2). 
  \qe
 Therefore, the~\mtg\ is a~central extension of~$T$ by~$Z(\Tg)$ with
a~factor system $m(\vt,\vt')$ resulting from multiplication of coset
representattives (cf.\ \refe{dec}, \cite{barut,kur})
 \eq 
 \tau(\vt\mid\vt)\tau(\vt'\mid\vt')=\tau(\vt\!+\!\vt'\mid\vt,\vt')=
\tau(\vz\mid\vt,\vt',-(\vt\!+\!\vt'))\tau(\vt\!+\!\vt'\mid\vt\!+\!\vt'),
\qe
 so
 \eq\label{facm}
 m(\vt,\vt')=\tau(\vz\mid\vt,\vt',-(\vt\!+\!\vt'))=
  \exp\left[-\half\ii e(\vt\tims\vt')\cdt\vH/c\hbar\right].
\qe

Therefore, (vector) representations of~$\Tg$ can be constructed as products
$\Gamma(z,\vt)\!=\!\Delta(z)\Lambda(\vt)$, where~$\Delta$ is
a~representation of the center~$Z$ and~$\Lambda$ is a~projective
representation of~$T$ with a~factor system \cite{alt,kur}
 \eq 
  \nu(\vt,\vt')=\Delta(m(\vt,\vt')).
 \qe
 Such a~representation for $\Delta(z)\!=\!z$ is provided by a~mapping 
introduced by Brown~\cite{brown}
 \eq\label{brownd} 
 \Lambda(\vt)=\exp(-\ii\pib^{\,c}\cdt\vt/\hbar).
 \qe
 Therefore, this projective representation realize only one possible choice
of~$\Delta$\@. Considerations of all (irreducible) representations allow us
to get deep insight into algebraic structure of \mtg's and physical
relevance of their representations. 

 All these above mentioned facts suggest that \mtg's can be considered as
central extensions of~$T$ by~$G$, where $T\!\simeq\!\integ^3$ is the translation
group and $G\!\simeq\!Z(\Tg)$ is a~group of factors, so $G\subset\uo$\@.
However, in the next section the central extension of~$\integ^n$ by \uo\
will be investigated. The first change ($\integ^3\to\integ^n$) will allow us
to investigate a~general case of $n$-dimensional crystal lattice, where the
second change ($G\to\uo$) is done for the sake of simplicity and clarity of
considerations. It will then occur which factors $\ee^{\ii\phi}\in\uo$ form
the center of $\Tg$.

\section{Central Extensions of $\integ^n$ by \uo} \label{ext}
 It is easy to notice that any sequence $\vt_1,\ldots,\vt_j,-(\vt_1\!+\!
\ldots\!+\!\vt_j)$ corresponds to a~loop `drawn' in a~crystal lattice (i.e.\
using lattice vectors), so it is a~special case of a~path
$(\vt_1,\ldots,\vt_l$) (cf.~\refe{zakd})\@. From the group-theoretical point
of view the set of all paths is a~free group~$F$ generated by all non-zero
lattice vectors $\vt\in T$\@. On the other hand, all loops form the
kernel~$R$ of a~homomorphism $M\colon F\ra T$, which simply `calculates' the
value of a~path in~$T$, i.e.
 \eq
   M(\vt_1,\ldots,\vt_l)=\vt_1+\ldots+\vt_l.
\qe
 Moreover, each path can be written as a~product (in the group~$F$) of
a~loop and a~one-element path~$(\vt)$, which is chosen as the representative
of the right-coset in the decomposition $F\!=\!\bigcup_{\ivt} R (\vt)$\@.
These facts indicate close relations of Zak's approach with the \mlm\ for
determination all nonequivalent extensions of given groups. This method can
be used to determine the second cohomology group~$H^2(T,G)$ in a~general
case, i.e.\ for a~given action of~$T$ on~$G$\@. In the particular case,
considered in this paper, one is interested in (nonequivalent) central
extensions so this action is trivial. A~detailed description of this method
can be found in the original works of Mac~Lane \cite{mlm1,mlm2} or in some
other books and review articles (see \cite{kur,KM} or \cite{mll} and
references quoted therein)\@. Many examples of its application can be found
elsewhere~\cite{mlx1,mlx2}\@. Among others it has been applied to thorough
investigation of \mtg's in two dimensions \cite{mly}\@.

The main idea of the \mlm\ consists in replacing an exact sequence
 \eq
\{1\}\longrightarrow G \longrightarrow \Tg \longrightarrow T
\longrightarrow \{0\}
\qe
 by the following one
 \eq
\{1\}\longrightarrow R \longrightarrow F \longrightarrow T
\longrightarrow \{0\}.
\qe
  These sequences are related with each other by a~family\footnote{Strictly 
speaking these mappings form a~group, denoted by $\Hom_F(R,G)$, with
the point-wise composition rule $(\phi+\phi')(r)=\phi(r)\phi'(r)$.} of the
so-called operator homomorphisms $\phi\colon R\to G$\@. 

 In the following subsections the \mlm\ is realized step by step in the
case~$T\!\simeq\!\integ^n$ and $G\!=\!\uo$\@. Lattice vectors~$\vt$ will be
hereafter replaced by $n$-tuples $\vk:=(k_1,\ldots,k_n)$, $k_j\in\integ$ due
to the obvious isomorphism
 \eq\label{isotk}
 (k_1,k_2,\ldots,k_n)=\vk \leftrightarrow \vt_{\ivk}=\sum_{j=1}^n k_j\va_j,
 \qe
 where $\{\va_j\}_{j=1,\ldots,n}$ is a~crystal basis for a~given lattice.

\subsection{Generators of $\integ^n$}
  The very first step in the procedure is a~choice of generators of the
group $\integ^n$, i.e.\ the translation group~$T$\@. In this paper the most
natural set of generators is used, i.e.
 \eq\label{geneT}
  A_T:=\{\va_1,\va_2,\ldots,\va_n\}
 \qe
 consists of $n$~linearly independent vectors, which form a~crystal basis.
As the generators of $\integ^n$ we choose $n$-tuples 
 \eq\label{geneA}
  \ve_j:=(0,\ldots,0,k_j=1,0,\ldots,0),\qquad
A:=\{\ve_1,\ve_2,\ldots,\ve_n\},  
 \qe
  related with vectors $\va_j$ by the isomprphism~\refe{isotk}\@. Now a~free
group~$F$ of rank~$n$ has to be introduced. The alphabet~$X$ of this group
consists of $n$~letters $x_j$, $1\!\le\!j\!\le\!n$, such that 
 \eq\label{alphX}
   M(x_j)=\ve_j.
 \qe
 This formula determines an epimorphism $M\colon F\ra\integ^n$, since each
$m$-letter word $f^{(m)}\in F$ can be written as
 \eq\label{word}
  f^{(m)}=\prod_{k=1}^m \xi_k^{\veps_k},\qquad \xi_k\in X,\;\veps_k=\pm1,
 \qe
 and 
 $$
  M(f^{(m)})\!=\!\sum_{k=1}^m \veps_k M(\xi_k). 
 $$
 In the further considerations the inverse of any word $f\in F$ will be
denoted as $\fb$ and, of course, it will be also applied to the letters
$x_j\in X$, so $x_j^{-1}\!=\!\xb_j$.

\subsection{Decomposition of $F$}
 The kernel $\Ker M\!:=R\!$ consits of such words $f\in F$ that
$M(f)\!=\!\vz\!:=\!(0,\ldots,0)$ and the group~$F$ can be decomposed into
{\em right} cosets with respect to~$R$\@; since these cosets are
counter-images of~$M$ then they can be labelled by $\vk\in\integ^n$, so  
 \eq
   F=\bigcup_{\ivk}R f_{\ivk},
 \qe
 where $M(f_{\ivk})\!=\!\vk$\@. One may choose any
representatives~$f_{\ivk}$, but in the presented procedure it is important
that a~set of representatives
 \eq
  S:=\{f_{\ivk}\mid \vk\in\integ^n\}
 \qe
 is the Schreier set. It means that (cf.\ \cite{mll,kur,KM})
 \eqaxl{Schcon}
  &1_F\in S,\qquad X\subset S;&\label{S1}\\
  &\Bigl(f^{(m)}\neq 1_f,\;f^{(m)}\in S\Bigr)
   \Longrightarrow \Bigl(f^{(m')}\in S,\quad\forall\;
   m'=0,1,\ldots,m\!-\!1,\Bigr)&\label{S2}
 \aqex
 where $f^{(m')}$ denotes the $m'$-letter initial subword of $f^{(m)}$, i.e.
(cf.~\refe{word})
 \eq\label{subword}
  f^{(m')}=\prod_{k=1}^{m'} \xi_k^{\veps_k};\quad f^{(0)}:=1_F.
 \qe
  The conditions \refe{Schcon} are satisfied by the following set
 \eq\label{Schdef}
  S:=\{x_1^{k_1}x_2^{k_2}\ldots x_n^{k_n}\mid \vk\in\integ^n\}.
 \qe
 This definition determines also a~mapping $\Psi\colon\integ^n\ra F$ such
that
 \eq\label{Psi}
 \Psi(\vk)=\Psi\Bigl(\sum_{j=1}^n k_j\ve_j\Bigr) = 
 x_1^{k_1}x_2^{k_2}\ldots x_n^{k_n}; \quad (M\circ\psi)(\vk)=\vk.
   =f_{\ivk}.
 \qe
 A~composition of mappings~$\Psi$ and~$M$ in the opposite order is the
so-called choice function $\beta:=\Psi\circ M$, which maps each $f\in F$
onto the corresponding coset representative~$f_{\ivk}$, where $\vk\!=\!M(f)$
and $f\in R\,f_{\ivk}$\@. It is evident that each word $f\in F$ can be
written as a~product of $q$~words from the Schreier set~\refe{Schdef}, i.e.
 \eq\label{qword}
   f^{(m)}=\prod_{l=1}^q x_1^{k_1^{(l)}}\ldots x_n^{k_n^{(l)}},\qquad
   \sum_{l=1}^q\sum_{j=1}^n |k_j^{(l)}|=m.
 \qe
 Using this form of $f^{(m)}$ the choice function~$\beta$ is defined as 
 \eq\label{beta}
  \beta\Bigl(f^{(m)}\Bigr)=
  x_1^{\kappa_1}x_2^{\kappa_2}\ldots x_n^{\kappa_n},\qquad 
  \kappa_j=\sum_{l=1}^q k_j^{(l)}.
 \qe

\subsection{Alphabet of $R$}
   The next very important step is to find all factors $\vrho\colon \integ^n
\tims\integ^n\ra R$ determined as
 \eq \label{rho}
   \vrho(\vk,\vk'):=f_{\ivk}f_{\ivk'}\fb_{\ivk+\ivk'},
 \qe
 which can be also written as (recall that~$M$ restricted to~$S$ is
a~bijection)
 \eq\label{rhobis}
  \vrho(s,s')=ss'\ovr{\beta(ss')},
   \quad s,s' \in S.
 \qe
 It is easy to notice that this factor system is normalized since
 $$
  \vrho(1_F,s)=\vrho(s,1_F)=1_F\qquad{\rm or}\qquad
  \vrho(\vz,\vk)=\vrho(\vk,\vz)=1_F.
 $$
 All factors $\vrho(s,s')$ will be necessary in the last step of
this procedure but now one needs only a~part of them. The Nielsen--Schreier
theorem (see, e.g., \cite{KM}) states that the alphabet~$Y$ of the free
group~$R\subset F$ consists of nontrivial factors obtained for $s'\in
X$ (i.e.\ for $\vk'\in A$, see Eq.~\refe{geneA})\@. It means that
 \eqax\label{alphYa}
   Y&:=&\{y=s\xi\ovr{\beta(s\xi)}\mid 
    s\in S,\;\xi\in X,\;y\ne 1_F\}\\
   &=&\{y=f_{\ivk}f_{\ivk'}\fb_{\ivk+\ivk'}\mid 
   \vk\in\integ^n,\;\vk'\in A,\;y\ne 1_F\}.\label{alphaYb}
 \aqex
 It is easy to notice that there are $(n\!-\!1)$ {\em types} of such factors
that describe transformations, which are necessary to find a~representative
for any word $f\in F$\@. It follows from the fact that the letters of the
alphabet~$Y$ are determined in the procedure when the letters~$x_j$,
$j\!=\!1,\ldots,n\!-\!1$ are moved form the end of a~word 
$sx_j\!=\!x_1^{k_1}\ldots x_d^{k_d}x_j$ to the `proper', i.e.\ the $j$-th,
position. Therefore, these letters are connected with $(n\!-\!1)$ cyclic
permutations 
 $$
  (12\ldots n),\: (23\ldots n), \ldots, (n\!-\!1\,n)
 $$
 (for a~given $j\!=\!1,2,\ldots,n\!-\!1$ the last $n\!+\!1\!-\!j$ symbols
$x_l^{k_l}$ in the word~$s$ have to permuted in the cyclic order to `catch'
the added~$x_j$)\@. The letters of the alphabet~$Y$ will be hereafter denoted
as $A_j^{(\ivk)}$, where $\vk\!:=\!(k_1,\ldots,k_n)$, and can be calculated 
as
 \eqa
  A_j^{(\ivk)}&=&x_1^{k_1}\ldots x_j^{k_j}\ldots x_n^{k_n}x_j
 \Bigl(x_1^{k_1}\ldots x_j^{k_j+1}\ldots x_n^{k_n}\Bigr)^{-1}\nn
 &=&x_1^{k_1}\ldots x_j^{k_j}\ldots x_n^{k_n}x_j
 \xb_n^{k_n}\ldots\xb_j^{k_j+1}\ldots\xb_1^{k_1}.
 \label{Adef}
\aqe
 It is important to notice that for a~given~$j$ not all $n$-tuples~\vk\
determine nontrivial factors~$\vrho$ (i.e.\ letters~$A_j^{(\ivk)}$)\@. If all
exponents~$k_l$ for~$l\!>\!j$ are equal to zero than the corresponding
factor $A_j^{(\ivk)}\!=\!0$\@. It means that for a~given~$j$ all $n$-tuples
$\vk\!=\!(k_1,\ldots,k_j,0,\ldots,0)$ have to be omitted. It will occur
later on that a~set of parameters labelleling (nonequivalent) extensions
depends only on numbers $k_{j+1},\ldots,k_n$ (for each
$j\!=\!1,2,\ldots,n\!-\!1$)\@.
   
Two remarks are in place. At first, it is evident that the alphabet~$Y$
contains more (much more, in the considered case) letters than the
alphabet~$X$\@. Secondly, one can compare obtained results with those for
a~finite translation group $T\!\simeq\!\bigotimes_{j=1}^n \integ_{N_j}$\@.
In this case $n$-tuples $(k_1,\ldots,k_l,0,\ldots,0)$ are omitted only for
$l\!<\!j$\@. For $l\!=\!j$ one has to consider all \vk's with
$k_j\!=\!N\!-\!1$, which lead to letters $A_j^{(\ivk)}$ of the following
form
 $$
  A_j^{(k_1,\ldots,k_{j-1},N_j-1,0,\ldots,0)}
   =x_1^{k_1}\ldots x_{j-1}^{k_{j-1}}x_j^{N_j}
 \xb_{j-1}^{k_{j-1}}\ldots\xb_1^{k_1}.
 $$
 These letters appear in the considerations due to the relations, for
finite cyclic groups~$\integ_{N_j}$ or letters~$x_j$,
 $$
   N_j\va_j=\vz\quad{\rm or}\quad \beta(x_j^{N_j})=1_F.
 $$
 However, in the above formulae a~more general form of Eq.~\refe{Adef} was
used
 \eq
  A_j^{(\ivk)}=x_1^{k_1}\ldots x_j^{k_j}\ldots x_n^{k_n}x_j
 \Bigl(\beta(x_1^{k_1}\ldots x_j^{k_j+1}\ldots x_n^{k_n})\Bigr)^{-1}.
 \label{Adeff}
  \qe
 It is easy to determine a~number of letters~$A_j^{(\ivk)}$ (for the finite
translation group~$T$) as
 $$
  \sum_{j=1}^n \left[\Bigl(\prod_{l=1}^n N_l\Bigr)-
\Bigl(\prod_{l=1}^{j-1}N_l\Bigr)(N_j-1)\right] 
 = (n-1)\Bigl(\prod_{l=1}^n N_l\Bigr)+1
 $$
 what agrees with the Nielsen-Schreier theorem, which reads
 \eq\label{NieSch}
  |Y|=(|X|-1)|T|+1.
 \qe

  Any element of the kernel~$R$ can be written using letters of the
alphabet~$Y$; however, in actual calculations the alphabet~$X$ is rather
used, so one has to `translate' obtained words into the proper alphabet. The
appropriate formula has also been given by the Nielsen--Schreier theorem.
A~given word $f^{(m)}\in R$ (see Eq.~\refe{word}) can be written as
 \eq\label{transl}
  f^{(m)}=\prod_{k=1}^m\beta(f^{(k-1)})\xi_k^{\veps_k}
   \ovr{\beta(f^{(k)})},
  \qe
 where subwords~$f^{(k)}$ are determined in~\refe{subword} (cf.\ also
definition \refe{alphYa})\@.

It is convenient to introduce another set of letters (anothe alphabet
in~$R$), which are simply products of letters~$A^{(\ivk)}_j$ defined in the
following way
 \eq\label{rbyA}
  r_j^{(\vk)}:=\cases{
  \prod_{l=0}^{k_j-1} A_j^{(k_1,\ldots,k_{j-1},l,k_{j+1},\ldots,k_n)}, 
    & for $k_j>0$; \cr & \cr
  1_F, & for $k_j=0$; \cr & \cr
 \prod_{l=-1}^{k_j}\Ab_j^{(k_1,\ldots,k_{j-1},l,k_{j+1},\ldots,k_n)}&~\cr
 \;=\Bigl(\prod_{l=k_j}^{-1} A_j^{(k_1,\ldots,k_{j-1},l,k_{j+1},\ldots,k_n)}
   \Bigr)^{-1}, &  for $k_j<0$.
  }
 \qe
 On the other hand
 \eq\label{Abyr}
  A_j^{(\ivk)}=\rb_j^{(\ivk)}\,
    r_j^{(\ivk+\ive_j)}.
 \qe
 Considerations concernig the number of letters~$r$ (also for finite 
translations groups) are identical as for the letters~$A$\@. It is easy to
check (see the appendix) that  
 \eq\label{rper}
  r_j^{(\vk)}=x_1^{k_1}\ldots x_{j-1}^{k_{j-1}}x_{j+1}^{k_{j+1}}\ldots
    x_n^{k_n}x_j^{k_j}\Bigl(\beta(x_1^{k_1}\ldots x_n^{k_n})\Bigr)^{-1}.
 \qe
 These letters can be generalized to all permutations $\sigma\in S_n$ of the
indices $1,2,\ldots,n$\@, which are defined by the following formula 
(cf.\ \refe{rper} and \refe{rcom})
 \eq\label{rsigma}
   r_\sigma^{(\ivk)}
   =x_{\sigma(1)}^{k_{\sigma(1)}}\ldots
x_{\sigma(n)}^{k_{\sigma(n)}}\Bigl(x_1^{k_1}\ldots x_n^{k_n}\Bigr)^{-1}.
 \qe
 Of course we have
 $$
  r_j^{(\ivk)}=r_{(j\,j+1\,\ldots,n)}^{(\ivk)}.
 $$

\subsection{Action of $F$ on $R$}
 Now we have to consider conjugation of letters $A_j^{(\ivk)}\in Y$ by
letters $x_q\in X$, i.e.\ automorphisms of~$R$, which are inner ones
in~$F$\@. Obtained words, obviously belonging to~$R$, can be expressed in
the alphabet~$Y$, but it is more convenient to use both sets: letters
$A_j^{(\ivk)}$ and~$r_\sigma^{(\ivk)}$ (note that the latter set is not an
alphabet; only letters $r_j^{(\ivk)}$ corresponding to the cyclic
permutations form an alphabet)\@. It can be shown that 
 \eq\label{xAxb}
  x_q A_j^{(\ivk)} \xb_q =
\cases{ r^{(k_1,\ldots,k_{q-1},1,0,\ldots,0)}_{(q\,q-1\,q-2\,\ldots,1)}
  A_j^{(\ivk+\ive_q)}
  \rb^{(k_1,\ldots,k_{q-1},1,0,\ldots,0)}_{(q\,q-1\,q-2\,\ldots,1)},  & 
  for $j\ge q$; \cr  & \cr
  r^{(k_1,\ldots,k_{q-1},1,0,\ldots,0)}_{(q\,q-1\,q-2\,\ldots,1)}
  A_j^{(\ivk+\ive_q)}
  \rb^{(k_1,\ldots,k_j+1,\ldots,k_{q-1},1,0,
    \ldots,0)}_{(q\,q-1\,q-2\,\ldots,1)}, &
  for $j<q$.}
 \qe
  The symbols~$r_\sigma^{(\ivk)}$ should be rewritten using
letters~$r_j^{(\ivk)}$ or~$A_j^{(\ivk)}$, however we will see later on that
this form is much better for further calculations.

\subsection{Operator homomorphisms}
 A~homomorphism $\phi\colon R\ra\uo$ is called operator homomorphism if the 
following diagram is commutative:
 \eq\label{diag}
  \begin{array}{r@{}ccc@{}r}
      & R & {{\Xi_f}\atop{\longrightarrow}} & R & \\[2pt]
 \phi & \downarrow &           & \downarrow & \phi\,, \\[2pt]
      & \uo & {{\Delta(M(f))}\atop{\longrightarrow}} & \uo &
  \end{array}
 \qe
 where $\Xi_f(r)\!=\!fr\fb$ and $\Delta\colon T\ra \Aut \uo$ is
a~homomorphism describing an action of~$T$ on \uo\@. We consider the trivial
action~$\Delta$, so the operator homomorphisms satisfy
condition
  \eq\label{ophom}
    \phi(r)=\phi(fr\fb),\quad r\in R,\; f\in F.
  \qe
 It suffices to check this condition only for the alphabets~$X$ and~$Y$,
since $\phi$ is a~homomorphism, so the relation~\refe{xAxb} will be used. To
simplify notation the image of a~letter~$A_j^{(\ivk)}$ will be hereafter
denoted by~$a_j^{(\ivk)}\!=\!\phi(A_j^{(\ivk)})$\@. Since the group of
factors~\uo\ is abelian then for $j\!\ge\!q$ one obtains
 \eq\label{ophomcond1}
 a_j^{(\ivk)} = a_j^{(\ivk+\ive_q)}.
 \qe
 It means that $a_j^{(\ivk)}$ depends only on entries~$k_q$ for $j\!<\!q$\@.
In the special case we have that $a_{n-1}^{(\ivk)}$ depends only on~$k_n$\@.
The definition~\refe{rbyA} gives 
 \eq\label{phir}
  \phi(r_j^{(\ivk)})= (a_j^{(\ivk)})^{k_j}.
 \qe
 It is easy to prove that
 \eq\label{rsigmabyr}
 r^{(k_1,\ldots,k_{q-1},1,0,\ldots,0)}_{(q\,q-1\,q-2\,\ldots,1)}=
  \prod_{l=1}^{q-1}r_l^{(k_1,\ldots,k_l,0,\ldots0,k_q=1,0,\ldots,0)}
  \qe
  so
 \eq
 \phi\Bigl(
  r^{(k_1,\ldots,k_{q-1},1,0,\ldots,0)}_{(q\,q-1\,q-2\,\ldots,1)}\Bigr)
 =
  \prod_{l=1}^{q-1}\Bigl(
  a_l^{(k_1,\ldots,k_l,0,\ldots0,k_q=1,0,\ldots,0)}\Bigr)^{k_l}.
  \qe
  Introducing a~parameter (recall that we are interested in $j<q$ now)
 \eq\label{alq}
   a_{j,q}:=a_j^{(0,\ldots,0,k_q=1,0,\ldots,0)}
 \qe
 and taking into account relation~\refe{ophomcond1} the above formula can be
written as
  \eq\label{phirsigma}
 \phi\Bigl(
  r^{(k_1,\ldots,k_{q-1},1,0,\ldots,0)}_{(q\,q-1\,q-2\,\ldots,1)}\Bigr)
 = \prod_{l=1}^{q-1} a_{l,q}^{k_l}.
  \qe
 Therefore, the second equation in~\refe{xAxb} leads to the following
condition
 \eq\label{ophomcond2}
  a_j^{(\ivk)}=
  a_j^{(k_1,\ldots,k_q+1,\ldots,k_n)}
  \Bigl(\prod_{l=1}^{q-1} a_{l,q}^{k_l}\Bigr)
 \Bigl(\prod_{l=1}^{q-1} a_{l,q}^{-k_l}\Bigr) a_{j,q}^{-1}
  \qe
  for $j<q$\@. Finally, from both conditions~\refe{ophomcond1}
and~\refe{ophomcond2} we obtain
  \eq\label{ophomcond}
   a_j^{(\ivk)}=\prod_{q=j+1}^n a_{j,q}^{k_q}.
  \qe

Therefore, each operator homomorphism~$\phi$ is determined by~${n\choose2}$
parameters~$a_{j,q}\in\uo$\@. These numbers can be replaced by corresponding
real parameters~$\alpha_{j,q}\in[0,2\pi)$ with
$a_{j,q}\!=\!\exp(\ii\alpha_{j,q})$\@.  The last ones can be arranged in
a~real upper-triangular $n$-dimensional matrix~$\mata'$
 \eq\label{matAp}
  \mata'_{j,q}:=\cases{\alpha_{j,q}, & for $j<q$; \cr &\cr
                      0, & otherwise.}
 \qe
 It can be shown that for finite lattices there are~$n$ additional
parameters corresponding to the periodic boundary conditions.

\subsection{Crossed homomorphisms}
 The operator homomorphisms form a~group denoted hereafter
as~$\Hom_F(R,T)$\@. The main result of Mac~Lane is the following theorem:
 \eq\label{mclth}
  H^2(T,\uo) = \Hom_F(R,\uo)/Z_\Delta^1(F,\uo)|_R,
 \qe
 where $Z_\Delta^1(F,\uo)|_R$ is a~group of one-cocycles (the so-called
crossed homomorphisms) $Z_\Delta^1(F,\uo)$, restricted to~$R$\@, i.e.
 $$
 Z_\Delta^1(F,\uo):=\{\gamma\colon F\to\uo\mid\gamma(ff')=\gamma(f)\,
  \Delta(M(f))[\gamma(f')]\}.
 $$
 In the case of central extensions (i.e.\ trivial action~$\Delta$) it means
that each mapping~$\gamma$ should be a~homomorphism and, therefore, it is
determined by values $\gamma(x_j)$, $x_j\in X$\@. 

Since all operator homomorphisms are determined by parameters~$a_{j,q}$
given by Eq.~\refe{alq}, then it is enough to calculate images of
homomorphism ($j\!<\!q$)
 $$\gamma\Bigl(A_j^{(0,\ldots,0,k_q=1,0,\ldots,0)}\Bigr)=
 \gamma(x_qx_j\xb_q\xb_j)=1\in\uo.
 $$
 Therefore, all operator homomorphisms determine nonequivalent central
extensions of $\integ^n$ by \uo\@.

\subsection{Factor systems}
  The last task is to find factor systems, which are determined as 
 \eq\label{facsys} 
  m'_{\phi}(\vk,\vk')=\phi(\vrho(\vk,\vk')),
 \qe
 where~$\vrho$ is defined in~\refe{rho}\@. It can be easily done if one
express all factors~$\vrho$ by symbols~$r_j^{(\ivk)}$\@.  
 $$
  \vrho(\vk,\vk')=(x_1^{k_1}\ldots x_n^{k_n})(x_1^{k'_1}\ldots x_n^{k'_n})
  (\xb_n^{k_n+k'_n}\ldots \xb_1^{k_1+k'_1}).
 $$
  Considering the product of the first two representatives~$s\in S$ in the
above formula it is convenient to introduce $n$-tuples 
 $$
  \vk_l:=(k_1+k'_1,\ldots,k_l+k'_l,k_{l+1},\ldots,k_n);\qquad\vk_0:=\vk,
  \quad \vk_n=\vk+\vk'.
 $$
 Using this symbols we can write
 \eqa
\Bigl(\prod_{j=1}^n x_j^{k_j}\Bigr)
  \Bigl(\prod_{j=1}^n x_j^{k'_j}\Bigr)  
 &=&\rb_1^{(\ivk_0)} r_1^{(\ivk_1)}x^{k_1+k'_1}\Bigl(\prod_{j=2}^n 
   x_j^{k_{1,j}}\Bigr)\Bigl(\prod_{j=2}^n x_j^{k'_j}\Bigr) \nn
  &=& \Bigl(\prod_{j=1}^q \rb_j^{(\ivk_{j-1})} r_j^{(\ivk_j)}\Bigr)
  \Bigl(\prod_{j=1}^q x_j^{k_j+k'_j}\Bigr)
  \Bigl(\prod_{j=q+1}^n x_j^{k_{1,j}}\Bigr)
  \Bigl(\prod_{j=q+1}^n x_j^{k'_j}\Bigr)\nn
  &=& \Bigl(\prod_{j=1}^{n-1} \rb_j^{(\ivk_{j-1})} r_j^{(\ivk_j)}\Bigr)
  \Bigl(\prod_{j=1}^{n} x_j^{k_j+k'_j}\Bigr)
 \aqe
 Therefore
 \eq\label{rhores}
  \vrho(\vk,\vk')=\prod_{j=1}^{n-1} \rb_j^{(\ivk_{j-1})} r_j^{(\ivk_j)}
 \qe
 and
 \eq\label{mres}
  m'(\vk,\vk')=\prod_{j=1}^{n-1} \phi(\rb_j^{(\ivk_{j-1})})
  \phi(r_j^{(\ivk_j)})
  =\prod_{j=1}^{n-1} \prod_{q=j+1}^n a_{j,q}^{k_qk'_j}
  =\exp\Bigl\{\ii\sum_{j=1}^{n-1}\sum_{q=j+1}^n\alpha_{j,q}{k_qk'_j}
  \Bigr\}.
 \qe

For a~given factor system (a~given matrix $\mata'$) the central extension of
$\integ^n$ by \uo\ is a~set of pairs $[\exp(\ii\alpha),\vk]$, where
$\exp(\ii\alpha)\in\uo$ and $\vk\!=\!(k_1,k_2,\ldots,k_n)$ with the following
multiplication rule
 \eq\label{mulrul}
  [\exp(\ii\alpha),\vk][\exp(\ii\alpha'),\vk']
  =\Bigl[\exp\{\ii\alpha+\ii\alpha'
  +\ii\sum_{j=1}^{n-1}\sum_{q=j+1}^n\alpha_{j,q}{k_qk'_j}\},
  \vk+\vk'\Bigr].
 \qe

\subsection{Some remarks}
 Before discussing obtained results in the special case $n\!=\!3$ (the most
interesting in physical applications) it is possible to state some general
properties of the derived formulae. At first, it has to be stressed that, in
general, `nonequivalent' does not mean `nonisomorphic' and, therefore, some
of obtained groups are isomorphic. For example, if one defines an
automorphism of~\uo\ as
 $$
  \exp(\ii\alpha)\mapsto \exp(\ii\eta\alpha),\quad 
   \eta\in\real,\; \eta\neq0
 $$
 then factors determined by matrices~$\mata'$ and $\eta\mata'$ lead to
isomporphic groups due to a~mapping
 $$
  [\exp(\ii\alpha),\vk]\mapsto [\exp(\ii\eta\alpha),\vk].
 $$

Applying the formula~\refe{mulrul} we can calculate product of `pure'
translations, i.e.\ elements $[1,\vk]$\@. It is very interesting to find
this in the case of generators of $\integ^n$, i.e.\ $\ve_j$\@. To be more 
precise, we calculate a~product corresponding, under the
isomorphism~\refe{isotk}, to a~loop constructed from vectors $\va_j$
and $\va_q$ 
 $$
[1,-\ve_q][1,-\ve_j][1,\ve_q][1,\ve_j]
 =[\exp(\ii\alpha_{j,q}),-\ve_q-\ve_j][\exp(\ii\alpha_{j,q}),\ve_j+\ve_q]
 =[\exp(\ii\alpha_{j,q}),\vz].
 $$
 Therefore, the parameters of a~given extension correspond with factors,
which are gained after completing `primitive' loops. Going in the oposite
direction we obtain 
 $$
[1,\ve_j][1,\ve_q][1,-\ve_j][1,-\ve_q]=[\exp(-\ii\alpha_{j,q}),\vz],
 $$                                          
 i.e.\ the inverse of the previous factor.

One more very important fact has to be pointed: the \mcl\ method does not
provide us with the group of two-coboundaries $B^2(\integ^n,\uo)$ and,
therefore, we do not find trivial factor systems in an {\em explicite}\/
way. It means that choosing generators (the set~$A$) and the representatives
(the set~$S$) one obtains factor systems in a~form, which, sometimes, may be
inconvenient in the further applications. Hence, it may be necessary to find
also (usually, by direct calculations) at least one element of the group
$B^2(\integ^n,\uo)$.

Trivial factor systems (in the case of central extensions) are determined by
a~normalized (i.e.\ $\psi(\vz)\!=\!1$) mappings $\psi\colon\integ^n\ra\uo$
according to the following formula
 \eq\label{trivial}
  \theta_{\psi}(\vk,\vk'):=\psi(\vk)\psi(\vk')/\psi(\vk+\vk').
 \qe

 One can consider the following mapping\footnote{For a~finite gauge
group~$G$ it may be impossible to find such $b\in G$ that $b^2\!=\!a$ for
each $a\in G$.}  $\psi\colon \integ^n\ra\uo$:
 $$
  \psi(\vk)=\exp\Bigl\{{\ii\over2}
  \sum_{j=1}^{n-1}\sum_{q=j+1}^n\alpha_{j,q}k_qk_j\Bigr\}.
 $$
 This mapping defines a~trivial factor system 
 \eq\label{trivialB}
  \theta(\vk,\vk')=
\exp\Bigl\{-{\ii\over2}\sum_{j=1}^{n-1}\sum_{q=j+1}^n\alpha_{j,q}
  (k_qk'_j+k'_qk_j)\Bigr\}.
 \qe
 A~product of the factor system~\refe{mres} and the trivial
one~\refe{trivialB} determines an equivalent extension with the following
factor system
 $$
  m(\vk,\vk')=m'(\vk,\vk')\theta(\vk,\vk')
  =\exp\Bigl\{-{\ii\over2}\sum_{j=1}^{n-1} \sum_{q=j+1}^n\alpha_{j,q}
  (k_jk'_q-k_qk'_j)\Bigr\}.
 $$
 Introducing matrices (tensors):
 \eqax
 \mata&=&\mata'-(\mata')^{\rm T};\quad \mata_{j,q}=\cases{
  \alpha_{j,q}, & for $j<q$;\cr 0, & for $j=q$; \cr
  -\alpha_{q,j}, & for $j>q$}\\
 (\vk\otimes\vk')_{j,q}&=&k_j^{\,}k'_q
 \aqex
 this factor system can be written as
 \eq\label{mlast}
  m(\vk,\vk')=\exp\Bigl\{-{\ii\over2}\mata\cdot(\vk\otimes\vk')\Bigr\},
 \qe
 where
 $$
 A\cdt B:=\prod_{j,q=1}^n A_{j,q}B_{j,q}
 $$
 is a~scalar product of matrcies (a~transvection of tensors~\cite{gol}).

It is obvious that obtained factor systems are periodic with respect to the
parameters~$\alpha_{j,q}$ with identical periods~$4\pi$\@. Comparing the
obtained formula with Eq.~\refe{facm} one can notice that
$(\vk\otimes\vk')/2$ corresponds to $(\vt\tims\vt')/2$, i.e.\ the area of
a~triangle determined by vectors $\vt$ and $\vt'$, whereas the antisymmetric
tensor~$\mata$ corresponds (up to multiplicative constants) to the external
magnetic field~$\vH$\@. It will be more clear considering the case
$n\!=\!3$, when three-dimensional vectors will be associated with the
introduced tensors.

\subsection{Special case: $n=3$} \label{ext3}

The final formula~\refe{mlast} of the previous section can be written down
in an explicite way for $n\!=\!3$ in the following form
 $$
  m(\vk,\vk')=\exp\Bigl\{-{\ii\over2}\left(\begin{array}{ccc}
  0 & \alpha_{1,2} & \alpha_{1,3} \\
  -\alpha_{1,2} & 0 & \alpha_{2,3} \\
  -\alpha_{1,3} & -\alpha_{2,3} & 0
  \end{array}\right)
  \cdot
\left(  \begin{array}{ccc}
 k_1k'_1 & k_1k'_2 & k_1k'_3 \\
 k_2k'_1 & k_2k'_2 & k_2k'_3 \\
 k_3k'_1 & k_3k'_2 & k_3k'_3 
  \end{array}\right)
  \Bigr\}.
  $$
  For both tensors one can find corresponding (three-dimensional)
vectors~\vh\ and~\vs, respectively, using the symbol $\veps_{jqp}$ (i.e.\
the three-dimensional Ricci symbol~\cite{gol}), namely
 \eqas
  h_j&=&\half\sum_{q,p}\veps_{jqp}\alpha_{q,p};\\
  s_j&=&\half\sum_{q,p}\veps_{jqp}k_qk'_p,
 \saqe
 so 
 \eqas
  \vh&=&[\alpha_{2,3},-\alpha_{1,3},\alpha_{1,2}];\\
  \vs&=&\half[k_2k'_3-k_3k'_2,k_3k'_1-k_1k'_3,k_1k'_1-k_2k'_1]
 \saqe
  and
  \eq\label{vlast}
  m(\vk,\vk')=\exp\{-\ii\vh\cdt\vs\},
  \qe
  where $\vh\cdt\vs$ denotes a~scalar product of vectors. Comparing~$\vs$
with 
 $$
  \half(\vt_{\ivk}\tims\vt_{\ivk'})=s_1(\va_2\tims\va_3)
  +s_2(\va_3\tims\va_1)+s_3(\va_1\tims\va_2)
 $$
 one can see that~$\vs$ can be considered as a~covariant vector
corresponding with the area $\vS\!=\!\half(\vt_{\ivk}\tims\vt_{\ivk'})$\@. On the other hand~$\vh$
is a~contravariant vector written in the basis $\{\va_j\}$ as
 $$
  \vh=\sum_{j=1}^3h_j\va_j=\frac{eV}{c\hbar}\vH
 $$
 i.e.\ $h_j\!=\!(eV/c\hbar)H_j$, where~$H_j$ is a~component of the magnetic
field~$\vH$ in the direction~$\va_j$ and
$V\!=\!\va_1\cdt(\va_2\tims\va_3)$\@. Substituing these formulae into 
Eq.~\refe{vlast} one obtains
  $$
  m(\vk,\vk')=\exp\Bigl\{-\ii\frac{e}{c\hbar}\vH\cdt\vS\Bigr\}
  $$ what agrees with formula~\refe{facm}\@. Introducing $\omega\!=\!hc/e$
being the fluxon, i.e.\ the elementary quantum of magnetic flux 
(cf.\ \cite{azbel,zak1,AB,pesh}) and replacing $\vk$ and $\vk'$ by the
corresponding vectors $\vt_{\ivk}\equiv\vt$ and $\vt_{\ivk'}\equiv\vt'$
this formula may be writtens as
 $$
  m(\vt,\vt')=\exp\Bigl\{-2\pi\ii\frac{\vH\cdt\vS}{\omega}\Bigr\}
 $$
 what means that for a~given vectors
$\vt$ and $\vt'$ the corresponding factor $m(\vt,\vt')$ is trivial if the 
total magnetic flux through the triangle $\vt,\vt',-(\vt\!+\!\vt')$ is equal 
to integer number of fluxons and these factors are periodic in the magnetic
field with periods
 $$
  H_j=\frac{2}{V}\omega|\va_j|.
 $$
 If components of the magnetic field $\vH$ in the crystal basis $\{\va_j\}$
are rational numbers then number of different factors $m(\vt,\vt')$ is
finite. 
 
\section{Final remarks}
 The main result of this work is the relation~\refe{mlast} (and discussion
below it), which show that a~magnetic field can be interpreted as an
$(n\tims n)$ antisymmetric tensor and factor systems in $n$-dimensional
\mtg\ is determined by a~transvection of tensors. However, this fact was not
used in all its aspects, among others covariant and contravariant tensors
were not introduced and discussed in details. Nevertheless, some important
general conclusions can be formulated: (i) there is no possibility to
introduce \mtg\ in one dimension; (ii) a~(contravariant) tensor of rank 
$n\!-\!2$ corresponds to~$\mata$ (see, e.g., \cite{gol}); (iii) in the
special cases, $n\!=\!2,3$, it means that~$\mata$ can be represented by
a~scalar and a~vector, respectively. It seems that investigations of \mtg's
in four dimensions allow us to introduce the electromagnetic tensor~$F$,
providing that we conisder discrete time, and to consider commutativity of
time and space displacements. 

It is also important to compare the (equivalent) factor systems~\refe{mres}
and~\refe{mlast}\@. Considering a~pair of primitive vectors, namely
$\va_j,\va_q$ with $j\!<\!q$, we obtain
 $$
 m'(\va_j,\va_q)=1,\qquad m'(\va_q,\va_j)=\exp(\ii\alpha_{j,q}),
 $$
 whereas
 $$
 m(\va_j,\va_q)=\exp(-\ii\alpha_{j,q}/2),\qquad
 m(\va_q,\va_j)=\exp(\ii\alpha_{j,q}/2),
 $$
 so the first system leads to {\em a}-symmetric and the second one --- to
{\em anti}-symmetric factors. It is obvious that the second system leads to
operators~$\tau$, introduced by Zak, and projective representations,
considered by Brown, with the corresponding factors. It means that using
the other (equivalent) factor system (e.g.~\refe{mres}) the appropriate
formulae~\refe{zakd} and~\refe{brownd} should be changed --- it can be done
choosing the other, asymmetric, gauge, for example the Landau gauge
(cf.\ \cite{harper,geypop,WZ}).

This work ends with a~general formula for a~factor system in \mtg\@.
Obviously, the further considerations should lead us toward physics,
especially by investigations of the irreducible representations and the
periodic boundary conditions. 

\section*{Acknowledgement}
  The author would like to thank Professor T.~Lulek for his encouragement
and illuminating discussions. The author is grateful to Dr~M.~Thomas and
Dr~S.~Wa{\l}cerz for valuable suggestions in preparing the manuscript.
Professor M.~Bo\.zejko and Professor J.~Granowski's comments on relations
of the presented problem with the theta-function are greatly appreciated.

\par
\setcounter{section}{0}
\setcounter{subsection}{0}
\setcounter{equation}{0}
\def\thesection{Appendix:}
\def\theequation{A.\arabic{equation}}

\section[]{Relations between letters $A_j^{(\vk)}$ and $r_j^{(\vk)}$}
\label{rels}

Since in the following considerations only one element of $n$-tuple \vk\ is
relevant, namely~$k_j$ for a~given~$j$, therefore symbols $A_j^{(\vk)}$ and
$r_j^{(\vk)}$ will be in this section replaced by $A_j^{(k_j)}$ and
$r_j^{(k_j)}$, respectively. It means that in the presented formula other
entries of~\vk\ are fixed and irrelevant.

To check the relation~\refe{Abyr} one has to consider separately cases
$k_j\!<\!-1$, $k_j\!=\!-1$, $k_j\!=\!0$, and $k_j\!>\!0$\@. In the special 
cases, $k_j\!=\!-1,0$ one obtains
 \eqas
 \rb_j^{(-1)}\,r_j^{(0)} &=& A_j^{(-1)},\\
 \rb_j^{(0)}\,r_j^{(1)} &=& A_j^{(0)}, 
 \saqe
  since $r_j^{(0)}\!=\!1_F$\@. For~$k_j\!>\!0$ and $k_j\!<\!-1$ the following
expressions have to be written
 \eqas
 \rb_j^{(k_j)}r_j^{(k_j+1)} &=&
  \Bigl(\prod_{l=0}^{k_j-1}A_j^{(l)}\Bigr)^{-1} \prod_{l=0}^{k_j} A_j^{(l)}
   =\prod_{l=k_j-1}^{0}\Ab_j^{(l)} \prod_{l=0}^{k_j} A_j^{(l)}
   =A_j^{(k_j)}.\\
 \rb_j^{(k_j)}r_j^{(k_j+1)} &=& \prod_{l=k_j}^{-1} A_j^{(l)}
 \prod_{l=-1}^{k_j+1}\Ab_j^{(l)}=A_j^{(k_j)}.
\saqe

Verifying relation \refe{rper} one has to distinguish carefully the cases of
infinite and finite translation groups due to differences between
definitions \refe{Adef} and \refe{Adeff}\@. For $k_j\!\ne\!N_j\!-\!1$ both
cases are identical and the definitions of $A_j^{(\vk)}$ and $r_j^{(\vk)}$
give (for $k_j\!>\!0$)
 \eqas
 r_j^{(k_j)}&=&
 \Bigl(x_1^{k_1}\ldots x_j^{0}\ldots x_n^{k_n}x_j
 \xb_n^{k_n}\ldots\xb_j^{1}\ldots\xb_1^{k_1}\Bigr)
\Bigl(x_1^{k_1}\ldots x_j^{1}\ldots x_n^{k_n}x_j
 \xb_n^{k_n}\ldots\xb_j^{2}\ldots\xb_1^{k_1}\Bigr)\\
 &&\times \ldots\Bigl(x_1^{k_1}\ldots x_j^{k_j-1}\ldots x_n^{k_n}x_j
 \xb_n^{k_n}\ldots\xb_j^{k_j}\ldots\xb_1^{k_1}\Bigr)\\
 &=&x_1^{k_1}\ldots x_j^{0}\ldots x_n^{k_n}x_j^{k_j}\xb_n^{k_n}
   \ldots\xb_j^{k_j}\ldots\xb_1^{k_1}.
 \saqe
  The relation~\refe{rper} for $k_j\!=\!0$ is evident and for $k_j\!<\!0$
a~proof is analogous. It is easy to notice that for a~finite translation
group (i.e.\ for $\beta(x_j^{N_j})\!=\!1_F$), when the
definition~\refe{Adeff} has to be used, one obtains 
 $$
 r_j^{(N_j)}=
 x_1^{k_1}\ldots x_j^{0}\ldots x_n^{k_n}x_j^{N_j}
  \xb_n^{k_n}\ldots\xb_j^{0}\ldots\xb_1^{k_1}.
 $$
 Please note, that letters~$r_j^{(\ivk)}$ can be considered as an `ordering'
operators (or `cyclic' generalization of commutator) since
 \eq\label{rcom}
  x_1^{k_1}\ldots x_j^{k_j}\ldots x_n^{k_n} =  
 \rb_j^{(\ivk)}\Bigl(x_1^{k_1}\ldots x_j^{0}\ldots x_n^{k_n}x_j^{k_j}\Bigr)
 \qe
 and this formula confirms correspondence between the cyclic permutations
$(j\,j\!+\!1\,\ldots\,n)$ and letters $r_j^{(\ivk)}$ (so as well letters 
$A_j^{(\ivk)}$)\@.

The relation~\refe{Abyr} suggests calculation of products $\rb_j^{(k_j)}
r_j^{(k_j+l)}$\@. Using the identity~\refe{rper} it is easy to prove that
 $$
 \rb_j^{(k_j)}r_j^{(k_j+l)}= 
   x_1^{k_1}\ldots x_j^{k_j}\ldots x_n^{k_n} x_j^{l}
        \xb_n^{k_n} \ldots \xb_j^{k_j+l} \ldots \xb_1^{k_1}
        =A_{j,l}^{(k_j)},
 $$
 where a~symbol~$A_{j,l}^{(k_j)}$ is generalization
of~$A_{j}^{(k_j)}\!=\!A_{j,1}^{(k_j)}$\@.  

The above presented considerations confirms that different alphabets (with
the same number of letters) may be used to write down words in a~given free
group. However, the choice of the alphabet~$Y$ is an essential part of the
\mlm\ and is determined by the choice of generators (the set~$A$) and the
representatives (the Schreier set~$S$)\@. For example, one may add to
generators all (non-zero) multiplicities~$k_j\va_j$\@. According to the
Nielsen--Schreier formula~\refe{NieSch} this will lead to increase in number
of letters in the alphabet~$Y$\@. However, it seems that
letters~$r_j^{(\ivk)}$ cannot be obtained by changes in~$A$ and/or~$S$ sets.
On the other hand, these letters have been generalized to all permutations
in Eq.~\refe{rsigma}\@. It is well-known that cyclic permuations generate the
symmetric group~$S_n$\@. Despite this fact letters~$r_\sigma^{(\ivk)}$ are
not expressed by letters~$r_j^{(\ivk)}$ in an analoguos way. If it were than
one could have used only two type of letters,
namely~$r_{(1\,2\,\ldots\,n)}^{(\ivk)}$ and~$r_{(1\,2)}^{(\ivk)}$\@. For
example, taking $n\!=\!3$ and~$\sigma\!=\!(132)$ we have
 $$
 r_{(132)}^{(\ivk)}=x_3^{k_3}x_1^{k_1}x_2^{k_2}
   \xb_3^{k_3}\xb_2^{k_2}\xb_1^{k_1}.
 $$
 This element of the kernel~$R$ can be rewritten using
letters~$r_j^{(\ivk)}$ in two ways. The first is indirect and is based on
the `translation'~\refe{transl} and the relation~\refe{Abyr}\@. The second
method leads directly to letters~$r_j^{(\ivk)}$\@. Going from left to rigth 
one can add letters~$x_j^0$ in such a~way that a~permutation determined by
full cycle $(j\,j\!+\!1\,\ldots\,n)$ is obtained. Next the
relation~\refe{rcom} is apllied and the procedure goes on (after removing
unnecessary~$x_j^0$)\@. In the presented example one obtains
 \eqas
 r_{(132)}^{(\ivk)} &=& 
   x_3^{k_3}x_1^{k_1}x_2^{k_2}\xb_3^{k_3}\xb_2^{k_2}\xb_1^{k_1}\\
 &=& (x_2^0x_3^{k_3}x_1^{k_1})x_2^{k_2}\xb_3^{k_3}\xb_2^{k_2}\xb_1^{k_1}\\
 &=& r_1^{(k_1\,0\,k_3)}(x_1^{k_1}x_3^{k_3}x_2^{k_2})
   \xb_3^{k_3}\xb_2^{k_2}\xb_1^{k_1}\\
 &=& r_1^{(k_1,0,k_3)}\,r_2^{(k_1,k_2,k_3)}x_1^{k_1}x_2^{k_2}x_3^{k_3}
   \xb_3^{k_3}\xb_2^{k_2}\xb_1^{k_1}.
 \saqe
 Finally we have obtained
 $$
  r_{(132)}^{(\ivk)} =r_{(123)}^{(k_1,0,k_3)}\,r_{(23)}^{(k_1,k_2,k_3)}
 $$
 in spite of fact that $(132)=(123)(123)$.

\end{document}